\documentclass[a4paper]{jpconf}
\usepackage{amsmath}
\usepackage{cite}
\usepackage{booktabs}
\usepackage{graphicx}
\begin{document}
\title{Chiral symmetry aspects in the open charm sector}

\author{T Buchheim${}^{1,2}$, T Hilger${}^{3}$ and B K\"{a}mpfer${}^{1,2}$}

\address{${}^{1}$ Helmholtz-Zentrum Dresden-Rossendorf, PF 510119, D-01314 Dresden, Germany}
\address{${}^{2}$ Technische Universit\"{a}t Dresden, Institut f{\"u}r Theoretische Physik, D-01062 Dresden, Germany}
\address{${}^{3}$ University of Graz, Institute of Physics, NAWI Graz, A-8010 Graz, Austria}

\ead{t.buchheim@hzdr.de, thomas.hilger@uni-graz.at, kaempfer@hzdr.de}

\begin{abstract}
QCD sum rules serve as tools to investigate changing hadronic properties in a hot and/or dense nuclear medium.
The role of chiral symmetry breaking and restoration effects in a medium can be addressed also in the heavy-light meson sector.
Thus, we consider Weinberg sum rules which refer to chiral partner mesons composed of a light and a heavy quark.
\end{abstract}

\section{Introduction}

Within the standard model of particle physics, the current quark masses are generated by the Higgs mechanism.
Due to strong interaction process, the hadrons acquire masses by far exceeding these current masses in the $u$, $d$ sector.
Still, the masses of hadrons, in particular mesons, composed of a $u$ (or $d$) and a heavy quark (e.\,g.\ $c$) exceed the quark masses too.
While ab initio QCD evaluations allow for an access to the hadron spectrum \cite{wittig} somewhat indirect methods also relate hadron properties, encoded in spectral functions for instance, with QCD quantities most notable the QCD condensates characterizing the vacuum state.
In a strongly interacting environment, e.\,g.\ within nuclear matter, the latter ones are supposed to be modified \cite{Thomas:2006nx}.
As a consequence, the hadron spectral functions can experience induced modifications, too.
The celebrated QCD sum rules \cite{svz79} represent a formalized framework wherein this connection of hadron properties and QCD condensates can be quantified.

Among the infinite tower of condensates as expectation values of QCD operators, the chiral condensate $\langle \bar q q \rangle$ is a particularly important quantity, signaling dynamical chiral symmetry breaking (DCSB).
Here, we highlight some chiral symmetry aspects in the open charm meson sector.
It is the beam energy range of forthcoming experiments at FAIR, NICA and J-Park where the onset of charm production is expected.
Thus, charm degrees of freedom continue the previous focus on strangeness as probes of compressed nuclear matter and medium modifications of hadrons and as a tool for controlling our understanding of hadrons in themselves.

\section{QCD sum rules to study in-medium chiral effects}

A starting point of sum rule evaluations is the causal current-current correlator which reads in a medium $ \Pi(q) = \int d^4x \; e^{iqx} \langle\ \mathrm{T} \left[ j(x) j^\dagger(0) \right] \rangle_{T,n} $ \cite{hatsukoike}, where the current $j$ refers to a construction  representing the quantum numbers and essential structure features of the hadron species under consideration.
The symbol $\langle\cdots\rangle_{T,n}$ denotes Gibbs averaging 
at finite temperature ($T$) and/or baryon density ($n$); at zero temperature and density, it reduces to the vacuum expectation value.
From analyticity of the correlator and from Cauchy's formula one can derive an in-medium dispersion relation by relating the phenomenological spectral density $\mathrm{Im}\Pi(s)/\pi = \rho_{T,n}(s)$ at low energies with the correlator evaluated by Wilson's operator product expansion (OPE) in the deep Euclidean region:
\vspace{-1.0mm}
\begin{align}
	\int_{-\infty}^{+\infty} d\omega \frac{\rho_{T,n}(\omega,\vec q)}{\omega-q_0} \tanh\left(\frac{\omega}{2T}\right) = \sum_O C_O(q) \langle O \rangle_{T,n} 
	\label{eq:disprel}\\[-7.0mm]\nonumber
\end{align}
with the asymptotic expansion in QCD condensates $\langle O \rangle_{T,n}$ related to the non-perturbative, long-range effects parameterizing the complex QCD ground state.
The Wilson coefficients $C_O$ are complex valued functions computable in perturbation theory.
Equation~\eqref{eq:disprel} connects hadronic spectral properties encoded in $\rho_{T,n}$ to the QCD ground state via condensates.
Hence, changes of hadron properties (masses, mass splittings of (anti-)particles, widths) carry signals of the way the vacuum changes in a nuclear environment \cite{weise94} quantified by condensates, such as the chiral condensate which would vanish in a chiral restoration scenario.

\begin{figure}%
	\centering
	\hfill
	\includegraphics[width=0.45\columnwidth]{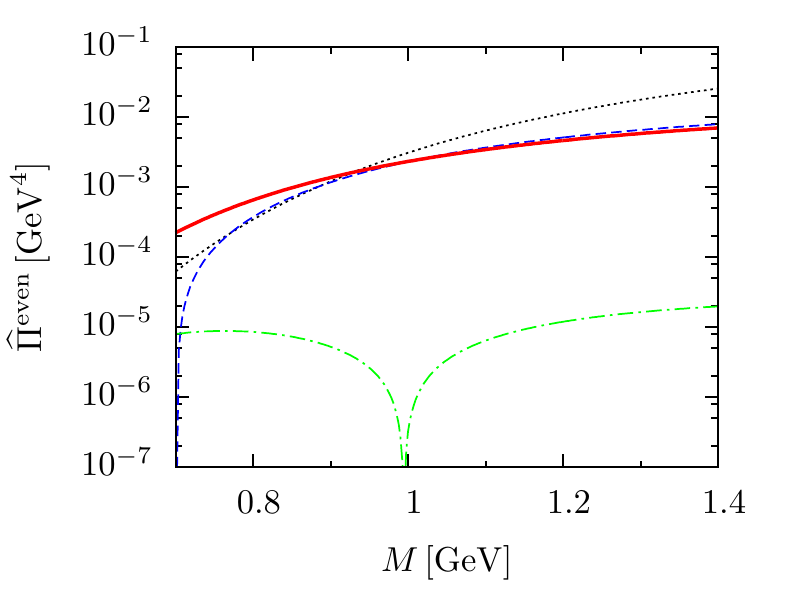}%
	\hfill
	\includegraphics[width=0.45\columnwidth]{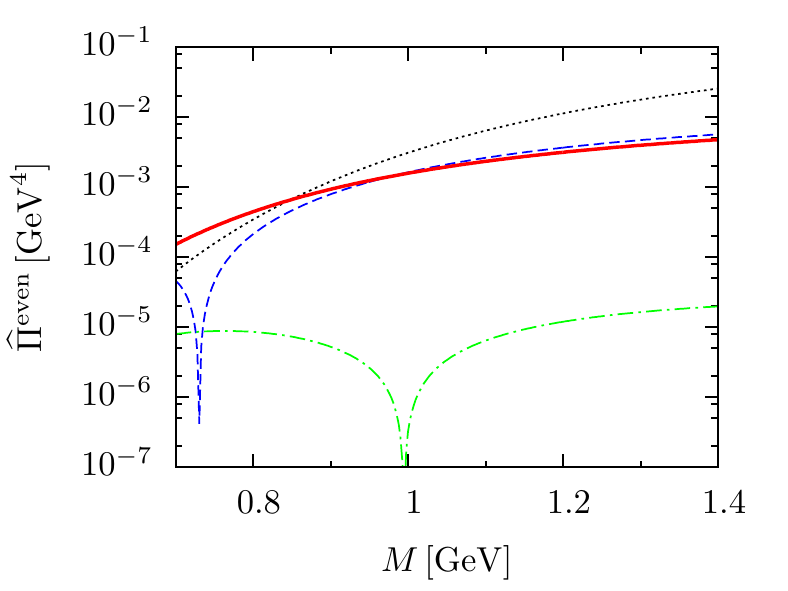}%
	\hfill
	\vspace{-2mm}
	\caption{Modulus of the contributions to the Borel transformed $D$ meson sum rule in vacuum (left panel) and at nuclear saturation density (right panel) \cite{buchheim15}. The chiral condensate (red solid) dominates at small Borel mass $M$ the non-perturbative contributions (blue dashed); the perturbative contribution (black dotted) and the four-quark condensate contribution (green dot-dashed) represent the largest and smallest contributions to the OPE, respectively.}%
	\label{fig:qqdom}%
	\vspace{-2mm}
\end{figure}

The chiral condensate can also be probed fairly well by mesons composed of a heavy ($Q$) and a light ($q$) quark.
For instance, in the $\rho$ and $\omega$ meson sum rules the chiral condensate contribution is suppressed numerically due to the light-quark mass, i.\,e.\ the renormalization group invariant combination $m_q \langle \bar q q \rangle$, whereas gluon and four-quark condensates are important instead \cite{rappwambachvanhees,Thomas:2007es}.
Such four-quark condensates have also turned out to be significant for nucleons \cite{Thomas:2006nk,thomas07}.
As a consequence, numerically relevant chiral symmetry effects enter the sum rule via chirally odd four-quark condensates \cite{hilger12}.
However, the chiral condensate has a sizable impact on spectral properties in the heavy-light meson sector, since the heavy quark acts as an amplification factor, i.\,e.\ the scale dependent combination $m_Q \langle \bar q q \rangle$ enters the sum rule \cite{Rapp:2011zz,zsch11,hilger10}.
The chiral condensate dominates in fact the contributions to the pseudo-scalar $D$ meson OPE in the region of interest (Borel window) in vacuum as well as at nuclear saturation density, see Fig.~\ref{fig:qqdom} \cite{hilger09,buchheim15}.
Thus, the $D$ meson sum rule is highly sensitive to chiral symmetry effects mediated by $\langle \bar q q \rangle$.

\section{Chiral symmetry transformations in the open charm sector}
\label{sec:ChiTrafo}

Chiral symmetry transformations act as flavour rotation on the left (L) and right (R) handed quark spinors separately: $ \phi_\mathrm{L,R} \longrightarrow e^{-i\Theta^a_\mathrm{L,R}t^a} \phi_\mathrm{L,R} \label{eq:chitrafo} $.
The quantities $\Theta^a_\mathrm{L,R}$ are the rotation para\-meters and $t^a$ denote the generators of the symmetry groups $SU(N_\mathrm{f})_\mathrm{L,R}$, where $N_\mathrm{f}$ is the number of involved quark flavours.
The mass term of the QCD Lagrangian breaks the symmetry.
However, if only light quark flavours are under consideration the approximation by massless quarks is justified and the QCD Lagrangian is invariant under such transformations.
Adding a heavy quark with mass $m_Q$ leaves the mass term of the QCD Lagrangian not invariant under chiral transformations, but one may restrict solely the flavour rotations on the light part of the flavour vector $\phi = \left( u, d, Q \right)^\mathrm{T} = \phi_\mathrm{L} + \phi_\mathrm{R}$ with $\phi_\mathrm{L,R} = \frac{1}{2}(1\mp\gamma_5)\phi$ by choosing those Gell-Mann matrices as generators which merely reduce to Pauli matrices forming the subgroup $SU(2)_\mathrm{L} \times SU(2)_\mathrm{R}$ of the full chiral symmetry group $SU(3)_\mathrm{L} \times SU(3)_\mathrm{R}$.
These re\-stricted transformations leave invariant the mass term $\bar \phi M \phi$ with $M=\mathrm{diag}(0,0,m_Q)$, but alter the chiral condensate, which can be transformed into its negative by a suitable choice of rotation parameters.
Thus, condensates that change under chiral transformations are dubbed chirally odd and serve as elements of order parameters of chiral symmetry.

One can show that chiral partner meson currents, such as pseudo-scalar and scalar $D$ meson currents, can be transformed into each other by an appropriate finite chiral transformation suggesting degenerate spectral properties of chiral partner mesons.
However, the experimentally observed heavy-light meson masses deviate strongly from one another.
This discrepancy signals DCSB, which is driven by order parameters, such as the chiral condensate.
The vacuum mass splitting of chiral partners (e.\,g.\ $P(0^-)$ states $\pi_{140}$, $D^0_{1865}$, $D^\pm_{s\,1986}$ vs.\ $S(0^+)$ states $f_{0\,400-550}$, $D^0_{2318}$, $D^\pm_{s\,2318}$ and $V(1^-)$ states $\rho_{775}$, $D^0_{2007}$, $D^\pm_{s\,2112}$ vs.\ $A(1^+)$ states $a_{1\,1230}$, $D^0_{2421}$, $D^\pm_{s\,2460}$) appears to be approximately 400$\,\mathrm{MeV}$ for light unflavoured as well as heavy-light mesons.
However, as the breaking pattern of light mesons is driven by four-quark condensates, whereas the one of open flavour mesons is driven by the chiral condensate, the modifications of spectra in both sectors might differ from each other.
Therefore, also the impact of four-quark condensates on open flavour mesons is investigated as a first step towards a comprehensive understanding of DCSB phenomenology \cite{hilgerbu12,buchheim14,buchheim14QCD14,buchheim14MESON14}.

\section{Weinberg sum rules}

Weinberg sum rules (WSR), known from current algebra, link moments of the difference of chiral partner spectra, $\Delta \rho = \rho^{(P)}_{T,n} - \rho^{(S)}_{T,n}$ or $\rho^{(V)}_{T,n} - \rho^{(A)}_{T,n}$, to order parameters of chiral symmetry, such as the pion decay constant, the chiral condensate and chirally odd four-quark condensates.
In this spirit, chirally odd condensates quantify the difference of chiral partner spectra.
Such difference sum rules attracted much attention as they can be utilized to investigate the medium behavior of vector and axial-vector spectral functions in the light meson sector.
The WSR of chiral partner mesons $\rho$ and $a_1$ have been studied extensively, e.\,g.\ to study chiral restoration driven by the order parameters of chiral symmetry \cite{hohlerrapp14}, to further restrict spectral functions containing more than one low lying resonance \cite{hohlerrapp12} as well as to check their compatibility with chiral mixing \cite{holthohlerrapp13} and hadronic models \cite{kwonweise10}.

According to the arguments of Sec.~\ref{sec:ChiTrafo},
one can set up Weinberg type sum rules as differences of chiral partner sum rules of heavy-light mesons in the $P$$-$$S$ as well as $V$$-$$A$ channels \cite{hilger11,Hilger:2009kn,Hilger:2010zb}. 
Evaluation of in-medium sum rules distinguishes even and odd parts of the OPE with respect to the meson energy.
We are able to extend the previous findings in the $P$$-$$S$ channel contributing solely to the even OPE $\int_{-\infty}^\infty d\omega \, \omega \, e^{-\omega^2/M^2} \tanh\left[\omega/(2T)\right] \Delta \rho(\omega)$ by (cf.\ \cite{hilger11,hilgerbu12} for details)
\vspace{-1.0mm}
\begin{align}
	e^{-m_Q^2/M^2} \Bigg[ -2m_Q \langle\bar q q\rangle + \left( \frac{m_Q^3}{2M^4} - \frac{m_Q}{M^2} \right)\langle\bar q g \sigma G q\rangle - \left( \frac{m_Q^3}{2M^4} - \frac{m_Q}{M^2} \right)\langle\Delta\rangle \Bigg]
		\label{eq:evenwsr}\\[-5.5mm]\nonumber
\end{align}
to mass dimension 6 by adding the chirally odd four-quark condensate contribution occurring in the odd OPE $\int_{-\infty}^\infty d\omega \, e^{-\omega^2/M^2} \tanh\left[\omega/(2T)\right] \Delta \rho(\omega)$ by
\vspace{-1.0mm}
\begin{align}
	\frac{2e^{-m_Q^2/M^2}m_Q}{M^4} g^2 \langle \bar q t^a q \sum_f q^\dagger_f t^a q_f \rangle \: .
	\label{eq:oddwsr}\\[-8mm]\nonumber
\end{align}
The expectation value $\langle\bar q g \sigma G q\rangle$ is the mixed quark-gluon condensate of mass dimension 5 and $\langle \Delta \rangle = \langle\bar q g \sigma G q\rangle - 8 \langle\bar q D_0^2 q\rangle$ \cite{hilger11} is a medium-specific condensate \cite{buchheim14QCD14} of mass dimension 5.

Pseudo-scalar $D$ and $B$ meson sum rule analyses in the nuclear medium \cite{hilger09} show that the odd OPE drives the mass splitting of anti-$D$($B$) and $D$($B$) mesons.
However, at zero nuclear density, particles and anti-particles are degenerate even at finite temperature involving a vanishing odd OPE \eqref{eq:oddwsr}.
Although, this requires a zero chirally odd four-quark condensate, chiral symmetry is not restored since the chirally odd condensates entering the even OPE \eqref{eq:evenwsr} do not vanish per se.
Therefore, chiral symmetry restoration does not emerge by vanishing of a single chirally odd condensate, but requires vanishing of all chirally odd condensates.

\section{Summary}

In summary, the role of chiral symmetry breaking and restoration in the open charm meson sector is addressed.
We consider Weinberg sum rules which refer to chiral partners.
Specific models, e.\,g.\ \cite{sasaki14}, should fulfil them, also evidencing that many chirally odd condensates beyond the chiral condensate $\langle \bar q q \rangle$ appear parameterizing the non-degeneracy of chiral partners prior to restoration.
Individual $D$ meson QCD sum rules for pseudo-scalar, scalar, vector and axial-vector channels depend on four-quark condensates, however, with overwhelming impact of the chiral condensate.

\section*{Acknowledgments}

This work is supported by BMBF 05P12CRGHE and FWF P25121-N27.

\section*{References}

\bibliographystyle{
									 iopart-num}
\bibliography{lit}

\end{document}